\documentstyle[11pt]{article}

\begin{document}
\begin{center}

{\LARGE Spin structure of the pion in a light-cone representation}

\renewcommand{\thefootnote}{\fnsymbol{footnote}}

\vspace{10mm}
{\large Bo-Qiang Ma\footnote{
        Fellow of Alexander von Humboldt Foundation, at present in
        Institut f\"ur Theoretische Physik der
        Universit\"at Frankfurt am Main, Postfach 11 19 32,
        D-6000 Frankfurt, Germany}}

\vspace{8mm}
{\large Institute of High Energy Physics, Academia Sinica, P.O.Box 918(4),
Beijing 100039, China }

\end{center}

\vspace{10mm}
\begin{minipage}[t] {105mm}
  {\large \bf Abstract } \\

\vspace{5mm}
The spin structure of the pion is discussed by transforming the
wave function for the pion in the naive quark model into a
light-cone representation. It is shown that there are higher helicity
($\lambda_{1}+\lambda_{2}=\pm1$) states in the full light-cone wave
function for the pion besides the ordinary helicity
($\lambda_{1}+\lambda_{2}=0$)
component wave functions as a consequence from the Melosh rotation
relating spin states in light-front dynamics and those in
instant-form dynamics. Some low energy properties of the pion, such
as the electromagnetic form factor, the charged mean square radius,
and the weak decay constant, could be interrelated in this
representation with reasonable parameters.
\end{minipage}

\vspace{24mm}
{To be published in Z.Phys.A}

\break

\noindent
{1. Introduction}

\noindent
The light-cone formalism [1,2] provides a convenient framework for
the relativistic description of hadrons in terms of quark and gluon
degrees of freedom. There have been many studies on the valence state
wave function for the pion in light-cone formalism [2-9] or in
light-front dynamics[10-12]. It will be shown in this paper that the
spin structure of the pion in light-cone formalism is quite different
from that in the SU(6) naive quark model in considering the effect
from the Melosh rotation [13,14] relating spin states in light-front
dynamics and those in instant-form dynamics. A natural consequence is
the presence of the higher helicity ($\lambda_{1}+\lambda_{2}=\pm1$)
components in the full light-cone wave function for the pion besides
the ordinary helicity
($\lambda_{1}+\lambda_{2}=0$)
components. In fact, the Melosh rotation has been applied to explain
the "'proton spin puzzle" [14] and the emergency of the
$\lambda_{1}+\lambda_{2}=\pm1$ components in the pion
has been also realized
[7]. The purpose of this paper is to explore the
explicit form of the light-cone wave function for the pion and the
consequences of taking into account the $\lambda_{1}+\lambda_{2}=\pm1$
components in the description of several low energy properties of the
pion. It will be shown that the electromagnetic form factor, the
charged mean square radius, and the weak decay constant could be
reproduced by the harmonic oscillator wave function in the light-cone
representation with very reasonable parameters by taking into account
the contributions from the higher helicity states.

\noindent
{2. Intuitive argument and present status}

\noindent
We first give an intuitive picture to explain why there are higher
helicity states in the full light-cone wave function for the pion.
When a composite system is transformed from one reference frame to
another frame, every constituent's spin will undergo a Wigner
rotation [15], and these spin rotations may be not necessarily the
same since the constituents may have different internal motions. In
consequence the sum of the constituent's spin is not Lorentz
invariant. Hence the pion, composed of two constituents with opposite
spin in the rest frame, may have $\lambda_{1}+\lambda_{2}=\pm1$ spin
states in the infinite momentum frame, where $\lambda_{1}$ and
$\lambda_{2}$ are the constituent's spins along the infinite momentum
frame; i.e., they are the helicities of the two constituents. We know
that the instant-form dynamics in the infinite momentum frame is
equivalent to light-front dynamics in an ordinary frame[16], thereby
the spin structure for a composite system in light-front dynamics is
quite different from that in the ordinary instant-form dynamics in
considering the effect from the Wigner rotation.

It has been argued [17,14] that the Melosh rotation [13] relating
constituent quark and current quark can be understood as a special
Wigner rotation relating spin states in instant-form dynamics and
those in light-front dynamics. The consequences from considering the
Melosh rotation in the description of the pion low energy
properties have been investigated in several papers [8,9,12]. In
ref.[8] some kinematics corrections to the ordinary helicity
($\lambda_{1}+\lambda_{2}=0$) component wave function were considered
but the higher helicity ($\lambda_{1}+\lambda_{2}=\pm1$) states were
unfortunately ignored. The pion wave function was
represented in ref.[12] in terms of Pauli matrices and the physical
implication is unclear. An explicit representation of the pion wave
function was given by Kisslinger and Jacob [9] in terms of light-cone
Dirac spinors in conjunction with a momentum space wave function
evaluated from a light-cone Bethe-Salpeter formalism. However, the
spin structure for the pion in the light-cone formalism, such as the
presence of the higher helicity states, still remains unexplained.

\noindent
{3. The light-cone wave function for the pion}

\noindent
{a) The spin wave function}

\noindent
We present in this paper an alternative light-cone representation of
the pion full wave function by transforming the ordinary
instant-form SU(6) quark model wave function for the pion into
light-front dynamics. We start from the rest frame
($\vec{q}_{1}+\vec{q}_{2}=0$) instant-form (T) spin wave function of
the pion,
\begin{equation}
\chi^{\pi}_{T}=(\chi_{1}^\uparrow\chi_{2}^\downarrow-
\chi_{2}^\uparrow\chi_{1}^\downarrow)/\sqrt{2},
\end{equation}
In which $\chi_{i}^{\uparrow,\downarrow}$ is the two-component Pauli
spinor and the two quarks have 4-momentum
$q_{1}^{\mu}=(q^{0},\vec{q})$ and $q_{2}^{\mu}=(q^{0},-\vec{q})$,
with $q^{0}=(m^{2}+\vec{q^{2}})^{1/2}$, respectively. The
instant-form spin states $|J,s>_{T}$ and the front form (F) spin
states $|J,\lambda\;>_{F}$ are related by a Wigner rotation $U^{J}$,
\begin{equation}
|J,\lambda>_{F}=\sum_{s} U^{J}_{s\lambda}|J,s>_{T},
\end{equation}
and this rotation is called as Melosh rotation for spin-1/2
particles. One should transform both sides of eq.(1) simultaneously
to get the light-cone spin wave function for the pion. For the left
side, i.e., the pion, the transformation is particularly simple since
the Wigner rotations are reduced to unity. For the right side, i.e.,
two spin-1/2 quarks, each particle instant-form and front-form spin
states are related by the Melosh transformation[14],
\begin{equation}
\begin{array}{clcr}
\chi^{\uparrow}(T)=w[(q^{+}+m)\chi^{\uparrow}(F)-
    q^{R}\chi^{\downarrow}(F)];\\
\chi^{\downarrow}(T)=w[(q^{+}+m)\chi^{\downarrow}(F)+
    q^{L}\chi^{\uparrow}(F)],\\
\end{array}
\end{equation}
where $w=[2q^{+}(q^{0}+m)]^{-1/2}$, $q^{R,L}=q^{1}\pm i\;q^{2}$, and
$q^{+}=q^{0}+q^{3}$. Then we get the light-cone (or front form ) spin
wave function for the pion,
\begin{equation}
\chi^{\pi}(x,\vec{k}_{\perp})=\sum_{\lambda_{1},\lambda_{1}}
C^{F}_{0}(x,\vec{k}_{\perp},\lambda_{1},\lambda_{2})
\chi_{1}^{\lambda_{1}}(F)\chi_{2}^{\lambda_{2}}(F),
\end{equation}
where the component coefficients
$C_{J=0}^{F}(x,\vec{k}_{\perp},\lambda_{1},\lambda_{2})$,
when expressed in terms of the instant-form momentum
$q^{\mu}=(q^{0},\vec{q})$, have the forms,
\begin{equation}
\begin{array}{clcr}
C^{F}_{0}(x,\vec{k}_{\perp},\uparrow,\downarrow)=w^{1}w^{2}
[(q_{1}^{+}+m)(q_{2}^{+}+m)-\vec{q}^{2}_{\perp}]/\sqrt{2};\\
C^{F}_{0}(x,\vec{k}_{\perp},\downarrow,\uparrow)=-w^{1}w^{2}
[(q_{1}^{+}+m)(q_{2}^{+}+m)-\vec{q}^{2}_{\perp}]/\sqrt{2};\\
C^{F}_{0}(x,\vec{k}_{\perp},\uparrow,\uparrow)=w^{1}w^{2}
[(q_{1}^{+}+m)q_{2}^{L}-(q_{2}^{+}+m)q_{1}^{L}]/\sqrt{2};\\
C^{F}_{0}(x,\vec{k}_{\perp},\downarrow,\downarrow)=w^{1}w^{2}
[(q_{1}^{+}+m)q_{2}^{R}-(q_{2}^{+}+m)q_{1}^{R}]/\sqrt{2};\\
\end{array}
\end{equation}
which satisfy the relation,
\begin{equation}
\sum_{\lambda_{1},\lambda_{2}}
C^{F}_{0}(x,\vec{k}_{\perp},\lambda_{1},\lambda_{2})^{*}
C^{F}_{0}(x,\vec{k}_{\perp},\lambda_{1},\lambda_{2})=1.
\end{equation}
One sees that there are also two higher helicity
($\lambda_{1}+\lambda_{2}=\pm1$) components in the expression of the
light-cone spin wave function of the pion besides the
ordinary helicity ($\lambda_{1}+\lambda_{2}=\pm1$) components.

\noindent
{b) The momentum space wave function}

\noindent
We still need to know the momentum space wave function.
Unfortunately, there is no exact solution of the Bethe-Salpeter
equation for the pion at present and in practice one often makes
approximation to evaluate the momentum space wave function. A
commonly used one for mesons is the harmonic oscillator wave function
\begin{equation}
\varphi(\vec{q}^{2})=A\;exp(-\vec{q}^{2}/2\beta^{2})
\end{equation}
which is a non-relativistic solution of the Bethe-Salpeter equation
in an instantaneous approximation in the rest frame for mesons[18].
We indicate that the relation between the instant-form momentum
$\vec{q}=(q^{3},\vec{q}_{\perp})$ and the light-cone momentum
$\underline{k}=(x,\vec{k}_{\perp})$ is by no means unique, and in
practice one
needs to construct models relating them. In this paper we adopt the
connection[10-12]:
\begin{equation}
\begin{array}{clcr}
x=(q^{0}+q^{3})/M;\\
\vec{k}_{\perp}=\vec{q}_{\perp}, \\
\end{array}
\end{equation}
in which $M$ satisfies
\begin{equation}
M^{2}=\frac{\vec{k}^{2}_{\perp}+m^{2}}{x(1-x)}.
\end{equation}
{}From eq.(8) we find,
\begin{equation}
\begin{array}{clcr}
q^{0}=[xM+(m^{2}+\vec{k}^{2}_{\perp})/xM]/2;\\
q^{3}=[xM-(m^{2}+\vec{k}^{2}_{\perp})/xM]/2,\\
\end{array}
\end{equation}
thus we obtain
\begin{equation}
\begin{array}{clcr}
q^{+}=xM;\\
2q^{+}(q^{0}+m)=(xM+m)^{2}+\vec{k}^{2}_{\perp}.\\
\end{array}
\end{equation}
We notice
\begin{equation}
\vec{q}^{2}=\frac{\vec{k}^{2}_{\perp}+m^{2}}{4x(1-x)}-m^{2}.
\end{equation}
As there are ambiguities in extending the non-relativistic form wave
function into a relativistic one, we find that there are three
possible prescriptions for the transformed light-cone momentum space
wave function in the literature:
\begin{enumerate}
\item
The Brodsky-Huang-Lepage (BHL) prescription [2]
\begin{equation}
\varphi_{BHL}(x,\vec{k}_{\perp})
=A\;exp[-\frac{m^{2}+\vec{k}^{2}_{\perp}}{8\beta^{2}x(1-x)}] ;
\end{equation}
\item
The Teren'ev-Karmanov (TK) prescription [10-11]
\begin{equation}
\varphi_{TK}(x,\vec{k}_{\perp})
=A\;\sqrt{\frac{1}{2x(1-x)}}
exp[-\frac{m^{2}+\vec{k}^{2}_{\perp}}{8\beta^{2}x(1-x)}] ,
\end{equation}
where the factor $\sqrt{1/2x(1-x)}$ arises from the jacobian relating
$d^{3}\vec{q}/q^{0}$ and $d^{2}\vec{k}_{\perp}dx $;
\item
The Chung-Coester-Polyzou (CCP) prescription [12]
\begin{equation}
\varphi_{CCP}(x,\vec{k}_{\perp})
=A\;\sqrt{\frac{M}{4x(1-x)}}
exp[-\frac{m^{2}+\vec{k}^{2}_{\perp}}{8\beta^{2}x(1-x)}] ,
\end{equation}
where the factor $\sqrt{M/4x(1-x)}$ arises from the jacobian relating
$d^{3}\vec{q}$ and $d^{2}\vec{k}_{\perp}dx$.
\end{enumerate}
The three prescriptions differ in the factor related with the
jacobian adopted: $J^{2}_{BHL}=1$, $J^{2}_{TK}=1/2x(1-x)$,
and $J^{2}_{CCP}=M/4x(1-x)$.

\noindent
{c) Parameter fixing}

\noindent
Thereby we obtain the light-cone wave function for the pion
\begin{equation}
\psi=\varphi\chi
\end{equation}
in which the parameters are the quark mass $m$, the harmonic scale
$\beta$ and the normalization constant A. The wave functions for the
pion in previous work[8-9,12] were shown to be successful in
reproducing some low energy properties of the pion, such as the
electromagnetic form factor, the charged mean square radius and the
weak decay constant. We exam also these items for the wave functions
in this paper. As the wave function, eq.[16], can be considered as a
light-cone version of the SU(6) quark model wave function
such as in ref.[18],
we expect that the parameters $m$ and $\beta$ be not so
much different from the values $m=330$ MeV and $\beta=220$ MeV in
ref.[18,3]. The parameters are adjusted to fit the constraints
adopted by Kisslinger and Jacob[9]:
\begin{enumerate}
\item
the normalization condition
\begin{equation}
\int [d^{2}\vec{k}_{\perp}dx/16\pi^{3}]\psi^{*}\psi=
\int [d^{2}\vec{k}_{\perp}dx/16\pi^{3}]\varphi^{*}\varphi=1,
\end{equation}
which is essentially a valence quark dominance assumption[8];
\item
the weak decay constant $f_{\pi}=93$ MeV defined [2] from
$\pi\rightarrow\mu\nu$ decay process by $<0|\overline{u}
\gamma^{+}(1-\gamma_{5})d|\pi>=-\sqrt{2}f_{\pi}p^{+}$, thus one
obtains
\begin{equation}
\int_{0}^{1}dx \int\frac{d^{2}\vec{k}_{\perp}}{16\pi^{3}}
\frac{(k^{+}_{1}+m)(k^{+}_{2}+m)-\vec{k}^{2}_{\perp}}
{\{[(k^{+}_{1}+m)^{2}+\vec{k}^{2}_{\perp}]
[(k^{+}_{2}+m)^{2}+\vec{k}^{2}_{\perp}]\}^{1/2}}\varphi=
\frac{f_{\pi}}{2\sqrt{3}};
\end{equation}
\item
the charged mean square radius $<r^{2}_{\pi}>=0.439$ fm$^{2}$ [19]
evaluated numerically from
$<r^{2}_{\pi}>=-6\partial F_{\pi}(Q^{2})/\partial Q^{2}$ at
$Q^{2}=0$.
\end{enumerate}
We thus obtain $m=330$ MeV, $\beta=290$ MeV for the BHL prescription;
$m=330$ MeV, $\beta=280$ MeV for the TK prescription;
and $m=330$ MeV, $\beta=270$ MeV for the CCP prescription. One sees,
in comparison with the results in ref.[8,12], that the values of the
parameters above are more close to those used
in the conventional SU(6) quark model harmonic oscillator wave
function[18,3].

\noindent
{4. The pion form factor}

\noindent
One advantage of light-front dynamics is that the Wigner rotation
relating spin states in different frames is unity under kinematic
Lorentz transformation, thereby the spin structure of hadrons are the
same in different frames related by kinematic Lorentz transformation.
The electromagnetic form factor can be calculated from the
Drell-Yan-West formula[20]
\begin{equation}
F(Q^{2})=\sum_{\lambda_{i}}\int_{0}^{1} dx \int
\frac{d^{2}\vec{k}_{\perp}}{16\pi^{3}}
\psi^{*}(x_{i},\vec{k}_{\perp i},\lambda_{i})
\psi^{}(x_{i},\vec{k'}_{\perp i},\lambda_{i}),
\end{equation}
where $\vec{k'}_{\perp i}=\vec{k}_{\perp i}-x_{i}\vec{q}_{\perp}
+\vec{q}_{\perp}$ for the struck quark,
$\vec{k'}_{\perp i}=\vec{k}_{\perp i}-x_{i}\vec{q}_{\perp}$ for the
spectator quarks, and the virtual photon momentum $q_{\mu}$ is
specified with $q^{+}=0$ to eliminate the Z-graph contributions[1,2,21].
Other choice of $q_{\mu}$ will cause contributions from Z-graphs, and
it should give the same result as that in the $q^{+}=0$ case if all
the graphs are taken into account[22].
We thus obtain
\begin{equation}
F(Q^{2})=\int_{0}^{1} dx \int
[d^{2}\vec{k}_{\perp}/16\pi^{3}]\cal M
\rm
\varphi^{*}(x,\vec{k}_{\perp})
\varphi(x,\vec{k'}_{\perp}),
\end{equation}
where
$\vec{k'}_{\perp}=\vec{k}_{\perp}+(1-x)\vec{q}_{\perp}$
is the internal quark transverse momentum of the struck pion in the center
of mass frame, and
\begin{equation}
\cal M
\rm
=\frac{(a_{1}a_{2}-\vec{k}_{\perp}^{2})(a'_{1}a'_{2}-\vec{k'}^{2}_{\perp})
+(a_{1}+a_{2})(a'_{1}+a'_{2})\vec{k}_{\perp}\cdot
\vec{k'}_{\perp}}
{[(a_{1}^{2}+\vec{k}_{\perp}^{2})(a_{2}^{2}+\vec{k}_{\perp}^{2})
(a'^{2}_{1}+\vec{k'}^{2}_{\perp})
(a'^{2}_{2}+\vec{k'}^{2}_{\perp})]^{1/2}},
\end{equation}
in which $a_{i}=k_{i}^{+}+m$ and $a'_{i}=k'^{+}_{i}+m$, is the
contribution from the Melosh rotation. The above calculation does not
(or less so severely) suffer from the flaws recognized
in ref.[7]
in evaluating the "'soft'' form
factor. One can easily find that our treatment of the Melosh rotation
is also different from that of CCP by comparing eq.(21) above with eq.(19)
in ref.[12]. This explains why our parameters differ to theirs. The
comparisons of the calculated form factor with the data at low
$Q^{2}$ are shown in fig.1. One sees, in combination with the three
constraints, that several low energy properties of the pion, such as
the electromagnetic form factor, the charged mean square radius, and
the weak decay constant, can be interrelated in the three
prescriptions with very reasonable parameters by taking into account
the contributions from the higher helicity states.

Fig.2 presents the calculated form factor at higher $Q^{2}$ with the
above wave functions in the constituent quark q\=q configuration. The
calculated form factors, i.e., the unlabeled curves, fall off with
$Q^{2}$ quickly by taking into account the contributions from higher
helicity states. If we ignore the higher helicity states as was done
in ref.[8], the calculated pion form factors, which should be
approximately twice the magnitude of the curves labeled
$\lambda_{1}+\lambda_{2}=0$, could be comparable in size to the data
at currently available $Q^{2}$ in the three prescriptions. Thus the
conclusion in ref.[8] might be altered if a different momentum space
wave function other than the one specified there was used. By
properly taking into account the contributions from the higher
helicity states, we seem to have arrived the same conclusion as that
in ref.[4,8] that the "'soft contributions'' to the form factor are
insufficient to explain the data at currently available large $Q^{2}$
and other QCD terms are necessary at high momentum transfer ($\geq$ a
few (GeV/c)$^{2}$), even if we adopt different prescriptions for the
pion light-cone wave function. Of course, the above conclusion is
dependent on the specific form of the momentum space wave function
and relies on the valence quark dominance assumption,
thus may be altered if a different momentum space wave function,
instead of eq.(7), is used or the valence dominance assumption, i.e.,
eq.(17),
is removed. Nevertheless, we believe the results in
fig.2 imply that the constituent quark model in valence quark
configuration could be valid at low energy scale even up to
$Q^{2}\approx 1$ (GeV/c)$^{2}$ and that at higher resolution (i.e.,
$Q^{2}\geq$ a few (GeV/c)$^{2}$) the contribution from considering
the internal substructure of the constituent quarks should be further
introduced in such picture like the valon model proposed by Hwa[25].

\noindent
{5. Remarks and summary}

\noindent
We indicate that our treatment of the Melosh rotation, though simply,
is different from those in previous investigations. The introduction
of the higher helicity states into the hadronic light-cone wave
functions may be able to shed some light on several problems
concerning the applicability of perturbative QCD in high momentum
transfer region. The higher helicity states are likely the sources
for the "'helicity non-conserving'' behaviors [26] observed in
pp$ ^{\uparrow}$ scattering [27] and in $\pi$N$\rightarrow\rho$N
process [28]. The Melosh rotation also has implications in the spin
content of hadrons. It has been shown in ref.[14] that the observed small
value of the integrated spin structure function for protons, i.e.,
the spin EMC data [29], could be naturally understood within the
naive quark model by taking into account the effect from Melosh
rotation based on the facts that deep inelastic process probes the
light-cone quarks other than the instant-form quarks [1-2,21,30], and
that the spin of the proton is the sum of the Melosh rotated
light-cone spin of the individual quarks other than simply the sum of
the light-cone spin of the quarks directly [31-32]. Thereby the
effects from the Melosh rotation need more careful considerations.

In summary, we showed, by using the pion as an example, that the spin
structure of hadrons in light-cone formalism is quite different from
that in the SU(6) naive quark model in considering the effect from the
Melosh rotation. One example is the existence of the higher helicity
states in the light-cone wave function for hadrons besides the
ordinary helicity states. It is shown that some low energy properties
of the pion, such as the electromagnetic form factor, the charged
mean square radius, and the weak decay constant, could be interrelated
by the harmonic oscillator wave function in the light-cone
representation with very reasonable parameters by taking into account
the contributions from the higher helicity states.

\noindent
{\bf Acknowledgement}

The author acknowledges many helpful discussions with Prof.T.Huang
and Prof.Q.-X.Shen. He also thanks Prof.H.J.Weber for drawing his
attention to the light-cone quark model and the Melosh rotation.
The hospitality and support from Prof.W.Greiner at the
        Institut f\"ur Theoretische Physik der
        Universit\"at Frankfurt are grateful.

\break

{\large \bf References}
\begin{enumerate}
\item Brodsky, S.J.: In: Lectures on lepton nucleon scattering and
      quantum chromodynamics. Jaffe, A.,   Ruelle, D.(eds.), p.255,
      Boston: Birkh\"auser 1982;

      Brodsky, S.J.,   Lepage, G.P.: In: Perturbative quantum
      chromodynamics. Mueller, A.H.(ed.), p.93,  Singapore:
      World Scientific 1989 and references therein
\item Lepage, G.P.   Brodsky, S.J.: Phys.Rev.D{\bf 22}, 2157(1980);

      Brodsky, S.J., Huang,T.,   Lepage, G.P.: In: Particle and
      fields.
      Capri, A.Z.,   Kamal, A.N.(eds.), p.143, New York: Plenum 1983;

      Lepage, G.P., Brodsky, S.J., Huang, T.,   Mackenzie, P.B.: ibid.,
      p.83
\item Isgur, N.,   Llewellyn Smith, C.H.: Phys.Rev.Lett.{\bf 52}, 1080(1984)
\item Jacob, O.C.,   Kisslinger, L.S.: Phys.Rev.Lett.{\bf 56}, 225(1986)
\item Dziembowski, Z.,   Mankiewicz, L.: Phys.Rev.Lett.{\bf 58}, 2175(1987);

      Dziembowski, Z.: In: Nuclear and particle physics on the light
      cone.
      Johnson, M.B.,   Kisslinger, L.S.(eds.), p.166, Singapore:
      World Scientific 1989
\item Huang, T.: Nucl.Phys.B(Proc.Suppl.){\bf 7}, 320(1989);

      Huang, T.,   Shen, Q.-X.: Z.Phys.C-Particles and Fields {\bf
      50}, 139(1991)
\item Isgur, N.,   Llewellyn Smith, C.H.: Nucl.Phys.B{\bf 317}, 526(1989);

      Phys.Lett.B{\bf 217}, 535(1989);

      see also, Isgur, N.: Nucl.Phys.A{\bf 497}, 229c(1989)
\item Dziembowski, Z.: Phys.Rev.D{\bf 37}, 778(1987). In this paper the term
      $b[\zeta-x(1-x)]$ in (8) should be $2\zeta^{2}x(1-x)$
\item Kisslinger, L.S.,   Jacob, O.C.: In:
      Nuclear and particle physics on the light cone.
      Johnson, M.B.,   Kisslinger, L.S.(eds.), p.322, Singapore:
      World Scientific 1989;

      Jacob, O.C.,   Kisslinger, L.S.: Phys.Lett.B{\bf 243}, 323(1990)
\item Terent'ev, M.V.: Yad.Fiz.{\bf 24}, 207(1976)
      [Sov.J.Nucl.Phys.{\bf 24}, 106(1976)]
\item Karmanov, V.A.: Nucl.Phys.B{\bf 166}, 378(1980)
\item Chung, P.L., Coester, F.,   Polyzou, W.N.:
      Phys.Lett.B{\bf 205}, 545(1988)
\item Melosh, H.J.: Phys.Rev.D{\bf 9}, 1095(1974)
\item Ma, B.-Q.: J.Phys.G{\bf 17}, L53(1991)
\item Wigner, E.: Ann.Math.{\bf 40}, 149(1939)
\item See, Bjorken, J.D., Kogut, J.B.,   Soper, D.: Phys.Rev.D{\bf 3},
      1382(1971)  and references therein
\item Kondratyuk, L.A.,   Terent'ev, M.V.: Yad.Fiz.{\bf 31}, 1087(1980)

      [Sov.J.Nucl.Phys.{\bf 31}, 561(1980)]
\item See, e.g., Elementary Particle Theory Group, Peking University:
      Acta Phys.Sin.{\bf 25}, 415(1976); Isgur, N.: In: The new aspects of
      subnuclear physics. Zichichi, A.(ed.), p.107, New York: Plenum 1980
      and references therein
\item Dally, E.B., Hauptman, J.M., Kubic, J., Stork, D.H., Watson,
      A.B., Guzik, Z., Nigmanov, T.S., Riabtsov, V.D., Tsyganov, E.N.,
      Vodopianov, A.S., Beretvas, A., Grigorian., A.,
      Tompkins, J.C., Toohig, T.E.,
      Wehmann, A.A., Poirier, J.A., Rey, C.A., Volk, J.T., Rapp, P.D.,
      Shepard, P.F.:
      Phys.Rev.Lett.{\bf 48}, 375(1982)
\item Drell, S.D.,   Yan, T.-M.: Phys.Rev.Lett.{\bf 24}, 181(1970);

      West, G.: Phys.Rev.Lett.{\bf 24}, 1206(1970)
\item Ma, B.-Q.: Ph.D. Dissertation, Peking University 1989;

      Ma, B.-Q.,   Sun, J.: J.Phys.G{\bf 16}, 823(1990);
      Int.J.Mod.Phys.A{\bf 6}, 345(1991);

      Ma, B.-Q.: Phys.Rev.C{\bf 43}, 2821(1991)
\item Sawicki, M.: Phys.Rev.D{\bf 46}, 474(1992)
\item Amendila, S.R., Badelek, B., Batignani, G., Beck, G.A.,
      Bedeschi, F., Bellamy, E.H., Bertolucci, E., Bettoni, D., Bilokon,
      H., Bologna, G., Bosisio, L., Bradaschia, C., Budinich, M., Codino,
      A., Counihan, M.J., Dell'orso, M., D'ettorre Piazzoli, B., Fabbri,
      F.L., Fidecaro, F., Foa, L., Focardi, E., Frank, S.G.F., Giazotto,
      A., Giorgi, M.A., Green, M.G., Harvey, J., Heath, G.P., Landon,
      M.P.J., Laurelli, P., Liello, F., Mannocchi, G., March, P.V.,
      Marrocchesi, P.S., Menasce, D., Menzione, A., Meroni, E., Milotti, E.,
      Moroni, L., Picchi, P., Ragusa, F., Ristori, L., Rolandi, L.,
      Saltmarsh, C.G., Saoucha, A., Satta, L., Scribano, A., Spillantini,
      P., Stefanini, A., Storey, D., Strong, J.A., Tenchini, R., Tonelli,
      G., Von Schlippe, W., Van Herwijnen, E., Zallo, A.:
      Phys.Lett.B{\bf 146}, 116(1984);
      Nucl.Phys.B{\bf 277}, 168(1986)
\item Bedek, C.J., Brown, C.N., Holmes, S.D., Kline, R.V., Ripkin,
      F.M., Raither, S., Sisterson, L.K., Browman, A., Hanson, K.M.,
      Larson, D., Silverman, A.:
      Phys.Rev.D{\bf 17}, 1693(1978) and references
      therein
\item Hwa, R.C.: Phys.Rev.D{\bf 22}, 759(1980);

      Hwa, R.C.,   Lam, C.S.: Phys.Rev.D{\bf 26}, 2338(1982);

      See, also, Zhu, W.,   Shen, J.G.: Phys.Rev.C{\bf 41}, 1674(1990)
\item Farrar, G.R.: Phys.Rev.Lett.{\bf 56}, 1643(1986)
\item Cameron, P.R., et al.: Phys.Rev.D{\bf 32}, 3070(1985)
\item Heppelmann, S., et al.: Phys.Rev.Lett.{\bf 55}, 1824(1985)
\item EMC,  Ashman, J., Badelek, B., Baum, G., Beaufays, J., Bee,
      C.P., Benchouk, C., Bird, I.G., Brown, S.C., Caputo, M.C., Cheung,
      H.W.K., Chima, J., Ciborowski, J., Clifft, R.W., Coignet, G., Combley,
      F., Court, G., D'agostini, G., Drees, J., D\"uren, M., Dyce, N.,
      Edwards, A.W., Edwards, M., Ernst, T., Ferrero, M.I., Francis, D.,
      Gabathuler, E., Gajewski, J., Gamet, R., V.Gibson, V., Gillies, J.,
      Grafstr\"om, P., Hamacher, K., Von Harrach, D., Hayman, P., Holt,
      J.R., Hughes, V.W., Jacholkowska, A., Jones, T., Kabuss, E.M.,
      Korzen, B., Kr\"uner, U., Kullander, S., Landgraf, U., Lanske, D.,
      Lettenstr\"om, F., Lindqvist, T., Loken, J., Matthews, M., Mizuno,
      Y., M\"onig, K., Montanet, F., Nassalski, J., Niinikoski, T., Norton,
      P.R., Oakham, G., Oppenheim, R.F., Osborne, A.M., Papavassiliou, V.,
      Pavel, N., Peroni, C., Peschel, H., Piegaia, R., Pietrzyk, B.,
      Pietrzyk., U., Povh, B., Renton, P., Rieubland, J.M., Rijllart, A.,
      Rith, K., Rondio, E., Ropelewski, L., Salmon, D., Sandacz, A.,
      Schr\"oder, T., Sch\"uler, K.P., Schultze, K., Shibata, T.-A., Sloan,
      T., Staiano, A., Stier, H., Stock, J., Taylor, G.N., Thompson, J.C.,
      Walcher, T., Wheeler, S., Williams, W.S.C., Wimpenny, S.J.,
      Windmolders, R., Womersley, W.J., Ziemons, K.:
      Phys.Lett.B{\bf 206}, 364(1988);

      Nucl.Phys.B{\bf 328}, 1(1989)
\item Drell, S.D., Levy, D.J.,  Yan, T.-M.: Phys.Rev.{\bf 187}, 2159(1969);

      Phys.Rev.D{\bf 1}, 1035(1970);

      Drell, S.D.,   Yan, T.-M.: Ann.Phys.{\bf 66}, 578(1971)
\item Coester, F.: Helv.Phys.Acta {\bf 38}, 7(1965);
      In: Constraint's theory and
      relativistic dynamics.  Longhi, G.,   Lusanna, L.(eds.),
      p.159,
      Singapore: World Scientific 1987; In: The three-body force in the
      three-nucleon system. Berman, B.L.,   Gibson, B.F.(eds.),
      p.472,
      New York: Springer 1986
\item Chung, P.L., Coester, F., Keister, B.D.,   Polyzou, W.N.:
      Phys.Rev.C{\bf 37}, 2000(1988);
      see, also, Keister, B.D.: In: Nuclear and particle physics on the
      light cone.
      Johnson, M.B.,   Kisslinger, L.S.(eds.), p.439, Singapore:
      World Scientific 1989
\end{enumerate}

\break

{\large \bf Figure captions}
\begin{enumerate}
\item Fig.1. Pion form factors calculated with the pion wave
      functions in the three prescriptions at low $Q^{2}$. The solid,
dashed, and dotted curves are the corresponding results from the BHL,
TK, and CCP wave functions with the parameters $m$ and $\beta$ being:
$m=330$ MeV and $\beta=290$ MeV for the BHL prescription;
$m=330$ MeV and $\beta=280$ MeV for the TK prescription;
and $m=330$ MeV and $\beta=270$ MeV for the CCP prescription. The
data are taken from ref.[23].
\item Fig.2. Pion form factors calculated with the pion wave
      functions in the three prescriptions at high $Q^{2}$. The
corresponding parameters for the curves are the same as those in
fig.1. The curves labeled "'$\lambda_{1}+\lambda_{2}=0$'' are the
contributions from the $\lambda_{1}+\lambda_{2}=0$ component wave
functions in the corresponding prescription. If the
$\lambda_{1}+\lambda_{2}=\pm 1$ components are ignored, the
calculated results should be approximately twice the magnitude of the
curves labeled "'$\lambda_{1}+\lambda_{2}=0$'' due to the normalization
condition. The curves labeled "'$\cal M
\rm=1$''
are obtained by turning
off the effect from the Melosh rotation. The unlabeled curves are the
contributions from the full (i.e., the $\lambda_{1}+\lambda_{2}=0$ and
$\lambda_{1}+\lambda_{2}=\pm 1$ components) wave functions. The data
are taken from ref.[24].
\end{enumerate}

\end{document}